# Janus MoSSe nanotubes on one-dimensional SWCNT-BNNT van der Waals heterostructures

Chunxia Yang[1,2 †], Qingyun Lin[3 †], Yuta Sato[4], Yanlin Gao[5], Yongjia Zheng[2], Tianyu Wang[2], Yicheng Ma[2], Wanyu Dai[1], Mina Maruyama[5], Susumu Okada[5], Kazu Suenaga[6], Shigeo Maruyama[1,2 *], Rong Xiang[2 *]

1 Department of Mechanical Engineering, The University of Tokyo, Tokyo 113-8656, Japan

2 State Key Laboratory of Fluid Power and Mechatronic Systems, School of Mechanical Engineering, Zhejiang University, Hangzhou 310027, China

3 Center of Electron Microscopy, State Key Laboratory of Silicon and Advanced Semiconductor Materials, School of Material Science and Engineering, Zhejiang University, Hangzhou 310027, China

4 Nanomaterials Research Institute, National Institute of Advanced Industrial Science and Technology (AIST), Tsukuba 305-8565, Japan

5 Graduate School of Pure and Applied Sciences, University of Tsukuba, Tsukuba 305-8571, Japan.

6 The Institute of Scientific and Industrial Research, Osaka University, 8-1 Mihogaoka, Ibaraki, Osaka 567-0047, Japan

maruyama@photon.t.u-tokyo.ac.jp
xiangrong@zju.edu.cn

## Abstract

2D Janus TMDC layers with broken mirror symmetry exhibit giant Rashba splitting and unique excitonic behavior. For their 1D counterparts, the Janus nanotubes possess curvature, which introduce an additional degree of freedom to break the structural symmetry. This could potentially enhance these effects or even give rise to novel properties. In addition, Janus MSSe nanotubes (M=W, Mo), with diameters surpassing 40 Å and Se positioned externally, consistently demonstrate lower energy states than their Janus monolayer counterparts. However, there have been limited studies on the preparation of Janus nanotubes, due to the synthesis challenge and limited sample quality. Here we first synthesized $MoS_2$ nanotubes based on SWCNT-BNNT heterostructure and then explored the growth of Janus MoSSe nanotubes from $MoS_2$ nanotubes with the assistance of $H_2$ plasma at room temperature. The successful formation of the Janus structure was confirmed via Raman spectroscopy, and microscopic morphology and elemental distribution of the grown samples were further characterized. The synthesis of Janus MoSSe nanotubes based on SWCNT-BNNT enables the further exploration of novel properties in Janus TMDC nanotubes.

## 1. Introduction

Two-dimensional (2D) Janus transition metal dichalcogenides (TMDCs) represent a promising class of quantum material, with different chalcogen atoms at the top and bottom of the transition metal layer. [1, 2] The out-of-plane mirror symmetry is broken in this unique atomic configuration and thus has an intrinsic electric field due to the electronegativity difference of two chalcogens. [3-5] Investigations to date indicate that many intriguing physical properties emerge in Janus TMDC monolayers, including Rashba spin-orbit coupling, [6, 7] second harmonic generation (SHG), [8, 9] piezoelectric polarization, [10, 11] and others. [2]

Their one-dimensional (1D) counterparts, Janus TMDC nanotubes possess curvature, which provides an additional degree of freedom to break the structural symmetry and potentially enhance these effects. [12] As revealed in the investigation by Hung *et al*., the SHG intensity in Janus MoSSe/$MoS_2$ heterobilayer can be improved by strain engineering. [13] Cai *et al*. also reported that the optimal power conversion efficiency of 1D MoSSe/$WSe_2$ heterostructure-based solar cell can reach up to 6.25%, which is considerably higher than that of 2D MoSSe/$WSe_2$ (1.94%).[14] Besides, the MSSe (M = W, Mo) nanotubes with a diameter of more than 40 Å (Se outside the tube wall) are energetically more stable than the corresponding Janus monolayer.[15] According to theoretical prediction, the strain energy of Janus MoSSe nanotube has a minimum at diameter around 80 Å.[16] The enhanced performance and excellent stability of Janus TMDC nanotubes makes them superior for further practical applications. However, the preparation of Janus TMDC materials remains challenging. Currently, experimental efforts in the synthesis of Janus TMDC materials are still in the exploratory stage and primarily focused 2D Janus monolayer, [10, 17-20] while little attention has been paid to the preparation of 1D Janus nanotubes with high crystallinity. [21]

Here, we report the fabrication and characterization of Janus MoSSe nanotubes (MoSSeNTs) base on 1D van der Waals heterostructures, which consist of single-walled carbon nanotubes (SWCNTs) and hexagonal boron nitride nanotubes (BNNTs). $MoS_2$ nanotubes ($MoS_2$NTs) were synthesized first on the SWCNT-BNNT templates and then converted to Janus MoSSeNTs at room temperature with the assistance of $H_2$ plasma. Raman spectroscopy confirms the formation of Janus MoSSe structure. The transmission electron microscope (TEM) characterization results indicate that the structure of formed monolayer Janus MoSSe nanotubes remains intact. The elemental distribution and the sulfur/molybdenum (S/Mo) ratio for single- and double-walled TMDC nanotubes provide evidence that only the exposed outer layer S atoms in $MoS_2$ nanotubes would be replaced by Se. Furthermore, the cross-sectional images of the synthesized 1D Janus heteronanotube illustrate a coaxial and Se-Mo-S sandwiched structure. Besides, it is not easy to convert $MoSe_2$ nanotubes into Janus nanotube structure, which is consistent with the theoretical calculation.

## 2. Results and discussions

Figure 1a schematically shows the fabrication process of Janus MoSSe nanotubes. The SWCNT-BNNT coaxial heterostructures,[22, 23] loaded on the ceramic washer, serve as the growth substrate for the duration of the synthesis process. First, $MoS_2$NTs were grown on SWCNT-BNNT coaxial nanotubes by low-pressure chemical vapor deposition (LPCVD) method. Subsequently, the SWCNT-BNNT-$MoS_2$NT sample with selenium (Se) particles were placed in a low-pressure chamber, where S atoms were selectively replaced by Se under the treatment of $H_2$ plasma. For nanotube samples at different stages, the color change in sample films can be observed by naked eyes, as illustrated in Figure 1b. SWCNT-BNNT is initially transparent but subsequently acquires a yellow hue following the addition of $MoS_2$. Thereafter, it displays a change in coloration to brown following exposure to $H_2$ plasma. The difference can also be discerned in optical absorption spectrum (Figure S1a).

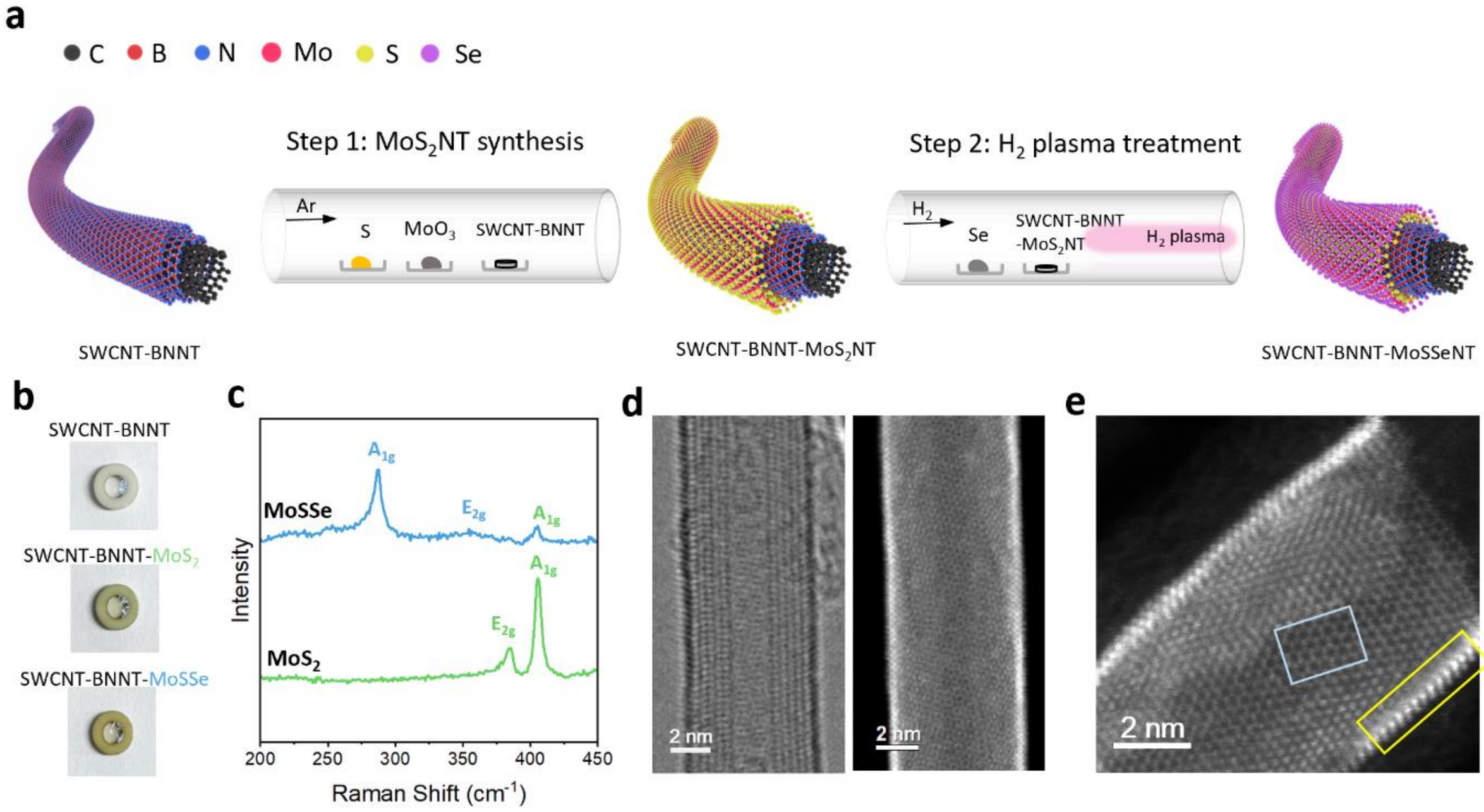

**Figure 1.** a) Schematic demonstration of the fabrication process of Janus MoSSe nanotubes based on 1D SWCNT-BNNT heterostructures. b) Optical image of the suspended SWCNT-BNNT, SWCNT-BNNT-$MoS_2$, and SWCNT-BNNT-MoSSe films on ceramic washers. c) Raman spectra of SWCNT-BNNT-$MoS_2$ and SWCNT-BNNT-MoSSe samples in the range from 200 to 450 cm$^{-1}$. d) TEM and HAADF-STEM images of SWCNT-BNNT-MoSSeNT. e) A magnified HAADF-STEM image of SWCNT-BNNT-MoSSeNT.

To confirm the formation of the Janus structure, Raman spectroscopy was employed to check the samples, both after $MoS_2$ synthesis and following the Janus conversion process. The overall Raman spectra are presented in Figure S1b, exhibiting characteristic peaks originating from CNT, BN, and $MoS_2$/MoSSe simultaneously within the range of 200 to 1800 cm$^{-1}$.[24, 25] For a magnified view of these Raman spectra within the 200 to 450 cm$^{-1}$ region, a clearer depiction is provided in Figure 1c. For the fabricated $MoS_2$ nanotube sample, characteristic in-plane $E_{2g}$ and out-of-plane $A_{1g}$ vibration mode can be found, centered at 386 cm$^{-1}$ and 406 cm$^{-1}$, respectively. [24] After the Janus MoSSe synthesis process, a new peak with strong intensity appeared at 287.3 cm$^{-1}$, this peak is corresponding to the out-of-plane S-Mo-Se vibration mode. At the same time, another distinct peak assigned for the in-plane Janus vibration mode at 354.5 cm$^{-1}$ appeared. [10] However, the $A_{1g}$ vibration mode from $MoS_2$ with significant lower intensity center at 406 cm$^{-1}$ can also be observed, indicating that majority (likely 70-80%) of but not all (reasons to discuss later) of $MoS_2$ was converted to the Janus MoSSe structure. In addition, the $A_{1g}$ and $E_{2g}$ for monolayer Janus MoSSe synthesized by similar method are located at 288 cm-

$^{1}$ and 355 cm$^{-1}$. [19] The peak positions of the MoSSe nanotubes are red-shifted compared to those of single-layer Janus MoSSe, suggesting a clear effect from the crystal curvature.

A typical TEM image of SWCNT-BNNT-MoSSeNT van der Waals heterostructure is shown in Figure 1d, where the coaxial structure is clearly distinguished. More TEM images are presented in Figure S2, revealing that the outermost layer of TMDC nanotubes is predominantly monolayers, with a minority of regions exhibiting the growth of bilayer TMDC nanotubes. Furthermore, as evidenced by the high-angle annular dark field scanning transmission electron microscopy (HAADF-STEM) characterization, the outmost MoSSe nanotubes are uniformly coated onto the SWCNT-BNNT template, with their crystalline structures remaining intact even after $H_2$ plasma treatment. A magnified HAADF-STEM image of a Janus MoSSe nanotube is shown in Figure 1e. It can be observed in the reference that, in $MoS_2$ nanotube samples, the Mo sites exhibit a significant higher brightness than the $S_2$ sites. [21] However, in Janus nanotube samples that have undergone Se substitution, the $S_2$ sites change into SeS and display an intensity that is comparable to that of the Mo sites, as indicated by the blue rectangle. Additionally, in the sidewall of the nanotube, as indicated by the yellow rectangle, the outer chalcogen atoms appear brighter than the inner ones, which can be attributed to the presence of Se and S atoms, respectively. [26]

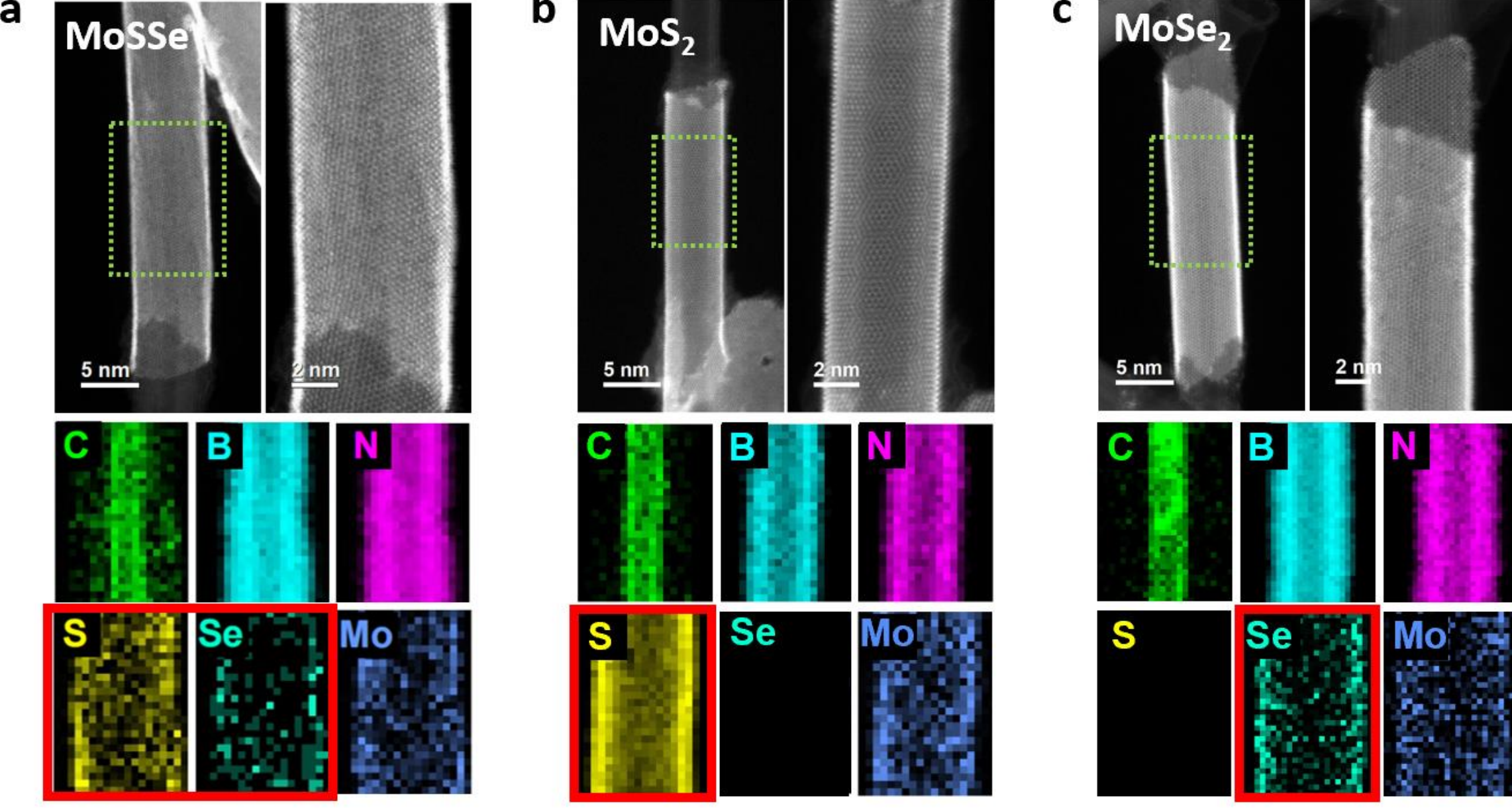

**Figure 2.** HAADF-STEM images and ELLS/EDS chemical maps for single-walled a) MoSSe, b) $MoS_2$, c) $MoSe_2$ nanotubes grown on SWCNT-BNNT coaxial heterostructures.

The distribution of elements in the fabricated SWCNT-BNNT-MoSSeNT is further verified by electron energy-loss spectroscopy (EELS) and energy-dispersive X-ray spectroscopy (EDS) mapping. The extensive elemental distribution across the sample is visually demonstrated in Figure S2g, which illustrates the large-area homogeneity. Figure 2a illustrates the distribution of elements within a single-walled Janus MoSSe nanotube. Notably, the innermost carbon (C) signal originates from the SWCNT, while the boron (B) and nitrogen (N) signals represent the BNNT in the middle. At the outermost layer, molybdenum (Mo), sulfur (S), and selenium (Se) are all present. In comparison to pure $MoS_2$ and $MoSe_2$ nanotubes grown on SWCNT-BNNT templates (Figure 2b, c), the S and Se signal intensities in Janus nanotubes are obviously lower than those in $MoS_2$ and $MoSe_2$ nanotubes.

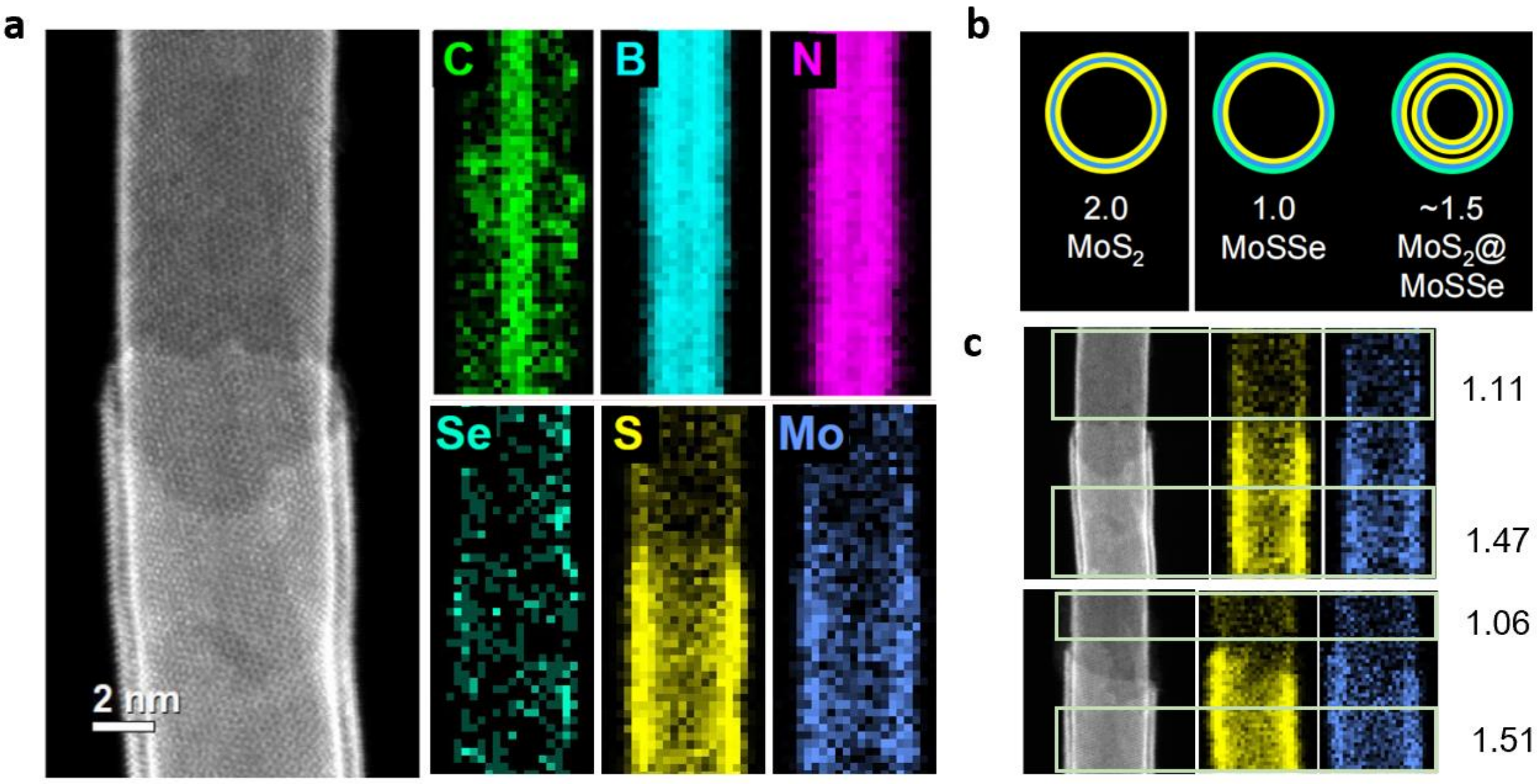

**Figure 3.** a) HAADF-STEM image and EELS/EDS chemical maps for single/double walled Janus MoSSe heteronanotubes. b) Schematic of $MoS_2$, MoSSe, and $MoS_2$@MoSSe nanotubes, and corresponding S/Mo ratio. c) HAADF-STEM images of synthesized single/double walled heteronanotubes, S and Mo elements distribution and the estimated S/Mo ratio.

In the synthesized Janus nanotube, featuring both single- and double-walled structures, the distribution of elements exhibits notable variations. As depicted in Figure 3a, within the upper single-walled segment, the presence of S, Se, and Mo elements is evident, with their signal intensities appearing comparable. Conversely, in the double-walled region, while all three elements (S, Se, and Mo) are still present, their signal intensities diverge significantly.

Specifically, transitioning from the monolayer to the bilayer area, the intensities of S and Mo undergo a marked enhancement, whereas the signal intensity of Se remains consistently low. This observation underscores the distinct elemental arrangements and interactions within the different wall configurations of the coaxial nanotube.

It is proposed that the disparate signal intensities of the three elements, S, Mo, and Se, within the single-walled and double-walled parts can be attributed to the selective replacement of outermost exposed S atoms on $MoS_2$ nanotubes. This hypothesis is corroborated by the estimation of the S/Mo ratio by EELS. As shown in Figure 3b, the S/Mo ration in $MoS_2$ nanotube is used as a reference and corrected to a value of 2.0. In monolayer Janus MoSSe nanotube, the S/Mo ratio should be decreased to 1, given that the outermost layer of S atoms has been replaced by Se atoms. In double-walled part, the S/Mo ratio should be approximately 1.5, as only the outermost S atoms have been substituted by Se, forming MoS2@MoSSe heteronanotube structures. In the synthesized samples, as shown in Figure 3c, the estimated S/Mo ration in the single-walled part is near 1 (1.11 and 1.06, respectively), and the value in the double-walled part is near 1.5 (1.47 and 1.51, respectively), which provides indirect evidence that only the exposed outermost layer of S atoms in $MoS_2$ nanotubes would be replaced by Se. This can also explain why some $MoS_2$ is still remaining even after a relatively long-time $H_2$ plasma treatment, as indicated by the Raman spectroscopy(Figure 1c previously). The elemental distributions of the sidewalls of Janus MoSSe nanotubes are also analyzed by EELS mapping (Figure 3S), but S and Se cannot be obviously distinguished from this (001) basal plane bird view. This is because side-walled crystals are geometrically highly curved in this 1D case, and the atoms at side walls are highly overlapped.

The cross-section characterization is an efficient approach for determining the structure of Janus structure. [10, 27, 28] In order to get a deeper understanding of the substitution of S atoms by Se within the Janus MoSSe nanotube, the heteronanotube cross-section sample was prepared by means of focused ion beam (FIB), and the characterization results are presented in Figure 4. The original nanotubes film on washer was first transferred onto a $SiO_2$/Si substrate, and it can be seen that the nanotubes are randomly distributed, as shown in Figure 4a. After being prepared into TEM cross-sectional samples, various morphologies of nanotubes could be observed, as depicted in Figure 4b. When the randomly distributed nanotubes are perpendicular to the cutting plane, a circular cross-section of the heteronanotubes becomes evident. Figure 4c illustrates the coaxial structure of SWCN-BNNT, featuring a SWCNT as the innermost nanotube, wrapped by several layers BNNT. Besides, the coaxial structure of SWCNT-BNNT-TMDCNT can be clearly observed in Figure 4d-e. The contrast levels of the two outermost layers are markedly disparate, indicative of the formation of TMDC nanotubes. On the basis of the S/Mo ratio, it can be concluded that the inner layer is still $MoS_2$ while the outer layer is Janus MoSSe. However, the MoSSe structure can currently only be partially observed (likely due to the initially discontinuous formation of MoS2 tube or damage to crystals in FIB preparation), as shown in Figure 4f, g. Meanwhile, when the FIB cutting plane is parallel to the heteronanotubes, an alternative perspective for observing the Se substitution is obtained, as depicted in Figure 4S.

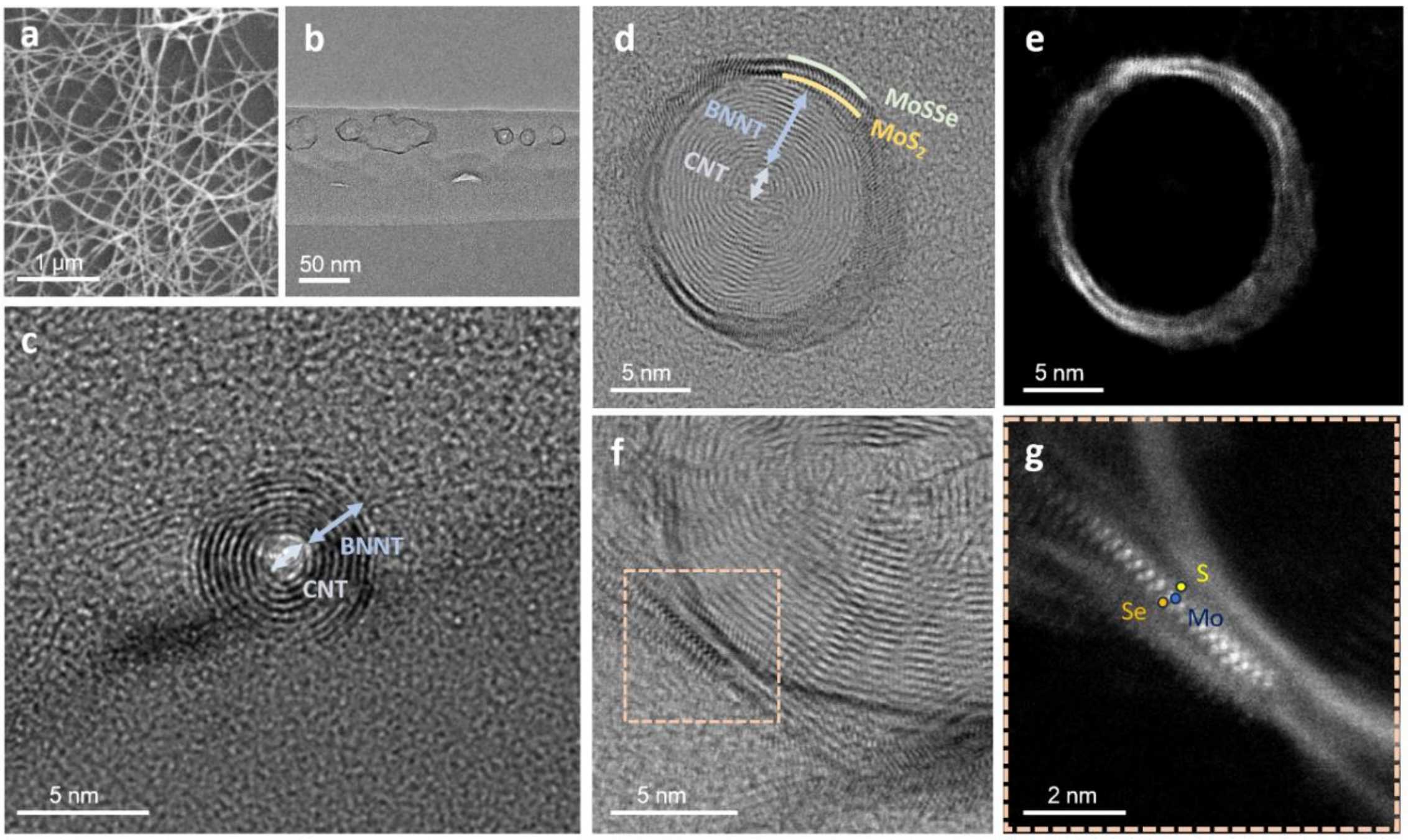

**Figure 4.** a) Scanning electron microscope image of randomly distributed heteronanotubes. b) TEM images of the sample cross-section. c) TEM images of SWCNT-BNNT coaxial structure. d, e) Cross-section annular bright field (ABF)-STEM and HAADF-STEM images of SWCNT-BNNT-$MoS_2$-MoSSe heteronanotubes. f, g) Cross-section images, showing the Janus MoSSe structure with S (yellow) inside and Se (orange) outside.

Finally, we compared the synthesis of 1D MoSSe and MoSeS Janus nanotubes. In case of 2D structures, selenizing monolayer $MoS_2$ or sulfurizing monolayer $MoSe_2$ gives identical planar Janus TMDCs monolayer.[18, 19] However, the resulted structure of the tubular Janus nanotubes differs significantly following selenization or sulfurization process. Given that only the outermost exposed chalcogen atoms are substituted, the presence of Se atoms on the outside of the nanotube (MoSSe) or S atoms on the exterior (MoSeS) can exert a considerable influence on the stability of the Janus nanotube structure. According the theoretical calculation as shown in Figure 5a, the total energy per atom of Janus nanotube with S outer is larger than that with Se outer, meaning that MoSSe Janus tubes are energetically more preferred. Further more, when the diameters exceed approximately 22 Å, Janus MoSSe nanotubes exhibit a lower strain energy than their corresponding Janus monolayer counterparts, suggested by the minus total energy shown in Figure 5a. Furthermore, previous calculations revealed the strain energy of Janus MSSe (M= Mo, W) nanotubes is smaller than Janus MSeS nanotubes, and attain their minimum strain energy at a diameter of approximately 80 Å for Janus MoSSe nanotubes.[15, 16, 29] We analyzed the diameters of Janus MoSSe nanotubes synthesized in this work (Figure 5b), the diameter is varying from 5 to 12 nm, with the majority falling within the range of 6 to 8 nm, which seems to be consistent with these previous predictions.

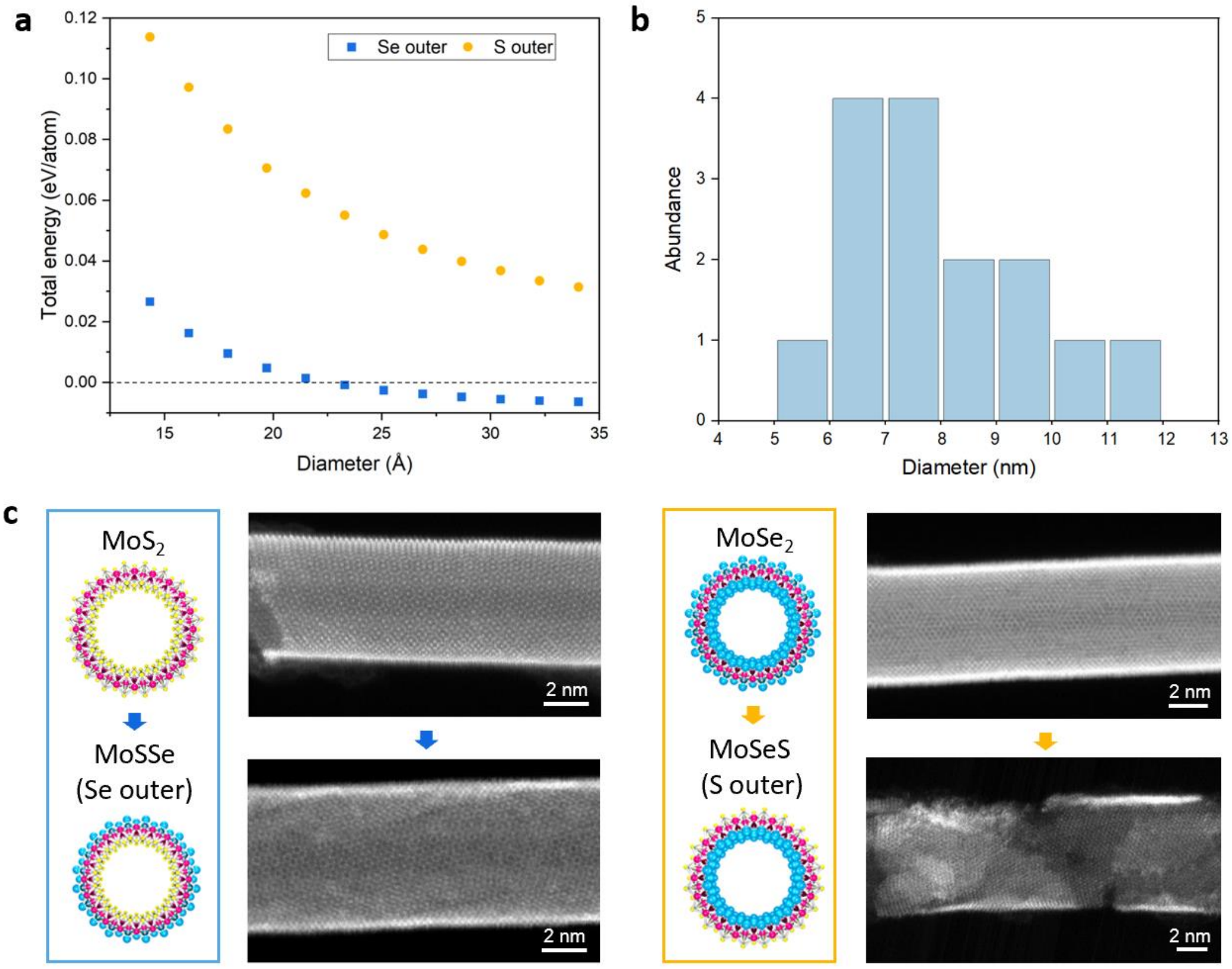

**Figure 5.** a) Total energy of Janus nanotubes with outermost walls of S and Se atoms as a function of diameters. The energy is measured to that of a monolayer Janus MoSSe. b) Histogram of the observed diameter of the Janus MoSSe nanotubes. c) Atomic model of transformation from $MoS_2$ nanotube to Janus MoSSe nanotube, and from $MoSe_2$ nanotube to Janus MoSeS nanotube, and corresponding HAADF-STEM images of $MoS_2$, MoSSe, $MoSe_2$, and MoSeS nanotubes.

At present, Trivedi and Guo *et al*. have successfully synthesized 2D Janus TMDC monolayers at room temperature through diverse experimental strategies, specifically selenization of $MoS_2$ monolayer and sulfuration of $MoSe_2$ monolayers.[18, 19] In these 2D cases, either selenizing MoS2 or sulfurizing MoSe2 appears similarly successful. The sulfurization process is believed to be even easier and facilitate the synthesis of Janus monolayers.[18] However, in the 1D experiments, the experimental results obtained from Janus nanotubes prepared through the selenization and sulfurization processes are entirely distinct. Although samples obtained through sulfurization also exhibited the characteristic Raman peak positions of a Janus, i.e. MoSeS, structure (Figure S1b, Figure S5), the conversion ratio is much lower the other approach. Furthermore, the microscopic characterization results indicate the high crystal quality of Janus MoSSe, whereas the Janus MoSeS structure is fragmented and unable to maintain the nanotube structure (Figure 5c upper and low). These observations clearly solidifies that, kinetically, formation of 1D MoSeS nanotube is slightly unfavorable, but energetically the structure of 1D MoSeS is significantly unstable. This is a distinct difference between 1D and 2D Janus crystals.

## 3. Conclusion

In summary, $MoS_2$ nanotubes grown on SWCNT-BNNT templates were converted into Janus MoSSe structures at the room temperature with the assistance of $H_2$ plasma. The formation of Janus MoSSe nanotubes was characterized by Raman spectroscopy. The microstructure and element distribution of SWCNT-BNNT-Janus MoSSe hetronanotubes are confirmed by TEM and ELLS/EDS mapping results. The TEM results proved the formation of monolayer and bilayer TMDC nanotubes on SWCNT-BNNT heterostructures. ELLS/EDS mapping and S/Mo ratio for single- and double-walled hetero-nanotubes indirectly prove that only the exposed outermost layer S atoms in $MoS_2$ nanotubes would be replaced by Se. From the cross-section images, coaxial nanotube structures can be distinguished and Se-Mo-S Janus structures can be partially found. Furthermore, it is advisable to prepare Janus MSSe nanotubes derived from $MS_2$ nanotubes, given the superior feasibility and stability of this preparation strategy. The successful fabrication of Janus MoSSe nanotubes paves the way for a deeper investigation into the novel properties exhibited by Janus TMDC nanotubes.

## 4. Experimental sections

*Preparation of SWCNT-BNNT growth template*: SWCNT film was first transferred onto a ceramic washer, and then place this washer loaded with SWCNT at the center of CVD furnace. 30 mg of ammonia borane ($H_3NBH_3$) was used as the BN precursor and positioned upstream of the SWCNT. In the synthesis process, BN precursor was heated to 70-90°C, while the SWCNT was heated to 1000-1100°C. Ar (with 3% $H_2$) was used as the carrier gas and the growth pressure maintained at 300 Pa. The BNNT growth is controlled for a duration of 1 hour. After the reaction, allow the furnace to cool naturally to room temperature before retrieving the SWCNT-BNNT sample, which will serve as the substrate for the subsequent growth of $MoS_2$ or $MoSe_2$ nanotubes.

*Synthesis of $MoS_2$ nanotubes based on SWCNT-BNNT*: $MoS_2$ nanotubes need to be prepared first for the fabrication of Janus MoSSe nanotubes. The $MoS_2$ nanotubes were synthesized using the LPCVD method, as shown in Figure 1. 30 mg of $MoO_3$ precursor was placed in a quartz boat at the first heating zone. SWCNT-BNNT template was place in another quartz boat at the second heating zone. The distance between $MoO_3$ and SWCNT-BNNT is around 8 cm. 6g of sulfur powder was placed upstream and heated to 130 °C by a heating belt. The furnace (heating zone 1 and heating zone 2) was ramped up with a rate of 30 °C $min^{-1}$ to 530 °C and held for 50 min before naturally cooling to room temperature. The Ar flow was maintained at 50 sccm throughout the experiment.

*Synthesis of Janus MoSSe nanotubes based on SWCNT-BNNT*: The as-grown SWCNT-BNNT-$MoS_2$ sample can be further used in the preparation of Janus MoSSe nanotubes. The schematic diagram of the related experiment is shown in Figure 1, where SWCNT-BNNT-$MoS_2$ was placed in the center of the quartz tube, Se particles were placed in the upstream region, and the downstream region was equipped with the home-build ICP setup. At the beginning of the experiment, it is necessary to evacuate the entire system to a base pressure of 2 Pa. Subsequently, 20 sccm $H_2$ was introduced as the carrier gas, maintaining the air pressure around 27 Pa. The whole plasma treatment process took about 20 min. At the end of the experiment, $H_2$ was turned off, Ar was refilled to atmospheric pressure, and the sample was removed.

*Synthesis of $MoSe_2$ nanotubes based on SWCNT-BNNT*: The synthesis of $MoSe_2$ nanotubes is different to $MoS_2$ nanotubes. A thick sapphire substrate has been designed to enable the SWCNT-BNNT film suspended on the $MoO_2$ precursor. The sapphire substrate, as shown in Figure S6a, has a counterbore in the center area, designed to accommodate the ceramic washer bearing the SWCNT-BNNT film. Figure S6b shows the schematic of $MoSe_2$ nanotubes growth. The Se particles are placed upstream of the molybdenum precursors and heated to 220 °C by a heating belt. This temperature ensures the sufficient evaporation of Se during the growth process. Besides, the $MoO_2$ and sapphire substrate are heated to 630 °C by the CVD furnace. A mixture of Ar and $H_2$ is used as the carrier gas with the flow rate at 50 sccm (Ar: 45 sccm; $H_2$: 5 sccm). During the experiment, the growth pressure is maintained at approximately 50 Pa.

*Synthesis of Janus MoSeS nanotubes based on SWCNT-BNNT*: The preparation process of Janus MoSeS nanotube was similar to that of Janus MoSSe. The SWCNT-BNNT-$MoSe_2$ heteronanotube sample and S powder were placed in the low-pressure chamber, and the experimental parameters, including the carrier gas, flow rate, and treatment time, were maintained at the same conditions as those utilized in the synthesis of Janus MoSSe.

*SWCNT-BNNT-MoSSe cross-section specimen preparation*: The ultra-thin S/TEM specimen of heteronanotube cross-section is prepared by combining the dual beam FIB precision manufacturing instrument (FEI Scios 2 Dual Beam) and Nano Mill system (Fischione 1040). Before the nanotube specimen was lifted out, a 2.0 μm thick carbon layer (0.5 μm e-beam deposition and 1.5 μm ion beam deposition) was deposited on the CNT surface to avoid Ga ion beam damage. After lift out, the specimen was thinned to 200-300 nm approximately with 1.0 μm carbon layer remained by using 16~5 kV Ga ion beam with different current. After the milling procedure in FIB, the lamella of nanotube cross-section is further milled by the low energy argon ions in the Nano Mill system, for completely cleaning the damaged surface caused by the FIB polishing and optimizing the specimen thickness to below 50 nm. The focused argon ion beam is range from 500 eV to 200 eV with beam current of 180 uA. Once the cross-section specimen preparation process by Nano Mill has finished, the samples were immediately stored in a glove box to protect them from being exposed to air for a long time.

*Characterizations*: Raman spectra of the samples were obtained via a backscattering configuration utilizing a Raman setup (Horiba, LabRAM Odyssey) with 532 nm excitation. A 100× microscope objective was employed to focus the laser beam and collect the scattered light. Absorption spectra were measured with a UV-vis-NIR spectrophotometer (Shimadzu, UV-3600i Plus). All measurements were carried out at room temperature. High resolution TEM images were acquired by using JEOL JEM F200 with an electron accelerating voltage of 200 kV. HAADF-STEM imaging combined with EELS/EDS chemical mapping was performed by using a JEOL JEM-2100F microscope equipped with double JEOL Delta correctors, a Gatan Quantum electron spectrometer, and double JEOL Centurio detectors at an accelerating voltage of 60 kV. The atomic-scale characterization of the heteronanotube cross-section with Janus structure is conducted in the spherical aberration corrected STEM (FEI Titan G2 80–200 ChemiSTEM), which is equipped with four Bruker energy dispersive X-ray spectrometers (super-EDS) for element analysis. For HAADF-STEM imaging, the inner and outer collection angles of the annular dark-field detector are set at 55 and 220 mrad, respectively.

*Density functional Theory calculation*: The calculations of the geometric structure of Janus nanotubes were based on density functional theory,[30, 31] employing the Simulation Tool for Atom Technology (STATE) program package.[32] The exchange-correlation potential between interacting electrons was expressed using the generalized gradient approximation (GGA) with the Perdew-Burke-Ernzerhof (PBE) functional.[33] Ultrasoft pseudopotentials generated by the Vanderbilt scheme were employed as the interaction between nuclei and electrons.[34] Valence wave functions and the deficit charge density were expanded in terms of plane wave basis sets with cutoff energies of 25 and 225 Ry, respectively. The atomic structures of the Janus nanotube were refined until the force exerted on each atom was minimized to below $1.33 \times 10^{-3}$ Hartree/bohr. For the lattice parameter aligned with the tube axis, we adopted a cell parameter of c = 0.3191 nm, derived from the optimum lattice parameter of an isolated MoSSe monolayer. The Brillouin zone integration was carried out with the use of 5 k points along the tube axis.

## Acknowledgement

C.Y. and Q.L. contributed equally to this work. Authors acknowledge Keigo Otsuka from the University of Tokyo and Sujuan Ding from Zhejiang University for the helpful discussion. This work was partially supported by JSPS KAKENHI (Grant Numbers JP23H00174, JP23H05443, JP21KK0087) and JST, CREST (Grant Number JPMJCR20B5), Japan. This work was also partially supported by the National Key R&D Program of China (2023YFE0101300, 2024YFA1409600) and Zhejiang province (2022R01001), China. TEM of this work was technically supported in part by "Advanced Research Infrastructure for Materials and Nanotechnology in Japan(ARIM)" of the MEXT, Japan.

# Supplementary Information

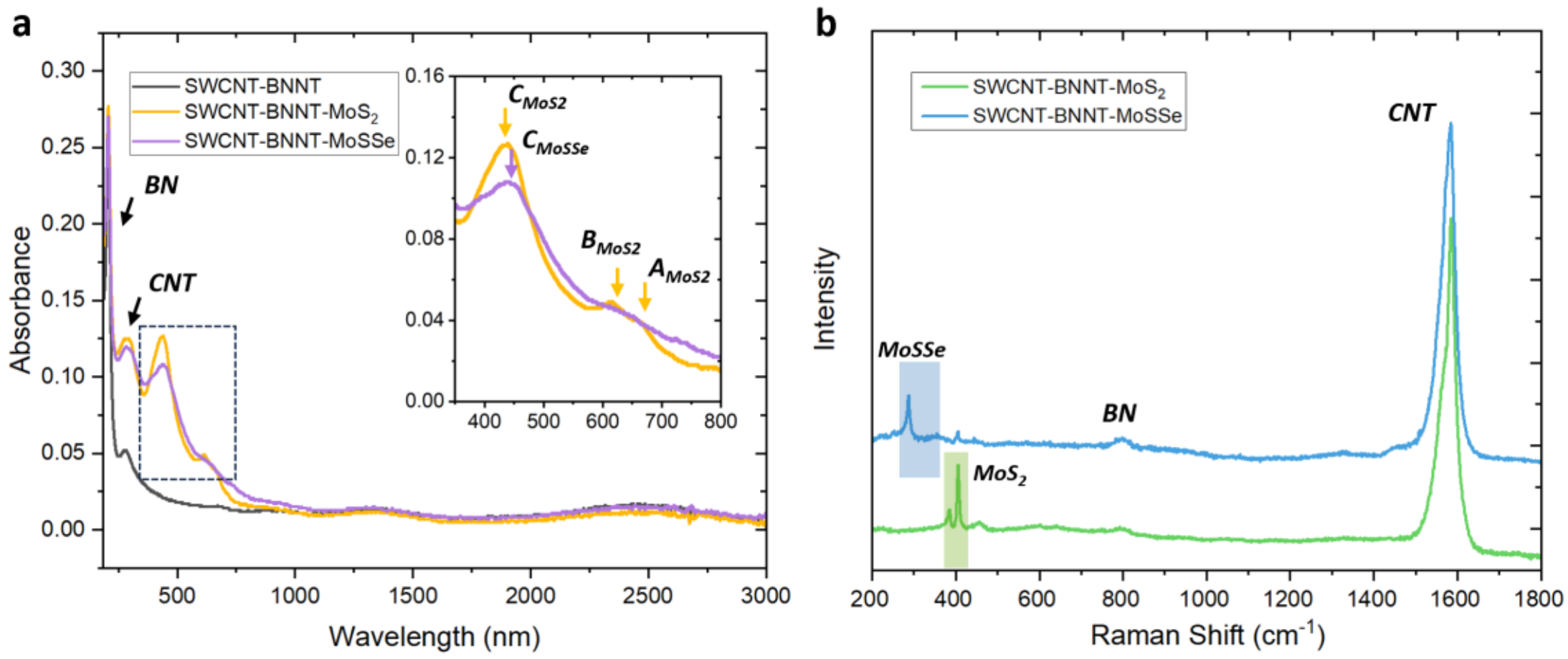

**Figure S1.** Optical spectra of heteronanotube samples at different stages. (a) Absorption spectra of pristine SWCNT-BNNT, SWCNT-BNNT-$MoS_2$, and SWCNT-BNNT-MoSSe. (b) Raman spectra of coaxial heteronantubes after $MoS_2$ synthesis and Janus MoSSe conversion processes.

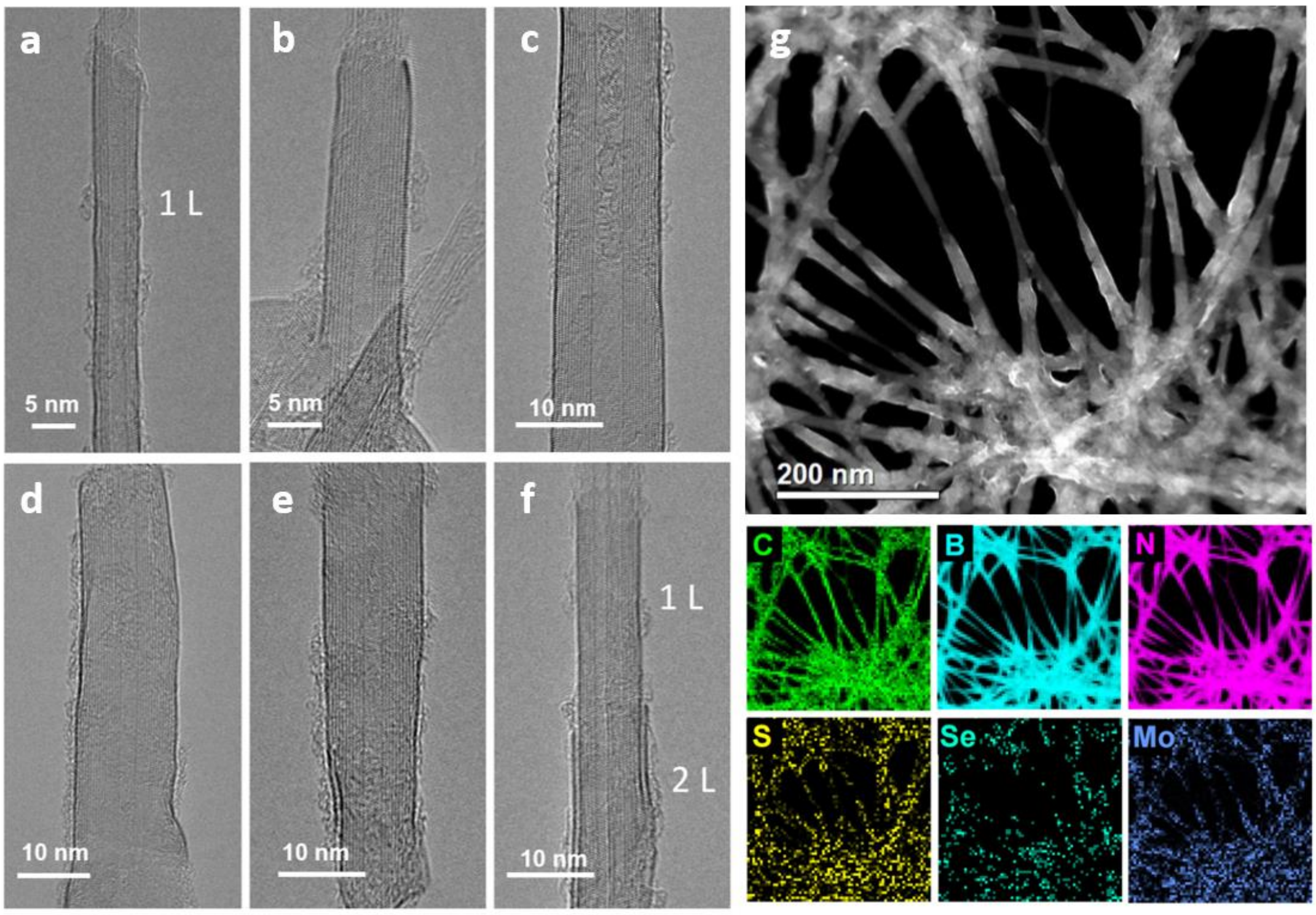

**Figure S2.** (a-f) TEM images of SWCNT-BNNT-TMDCNT, most TMDC nanotubes are monolayer. (g) HAADF-STEM image and EELS/EDS chemical maps of a relative extensive sample area. Mapping C/B/N/S/Mo by EELS and mapping Se by EDS.

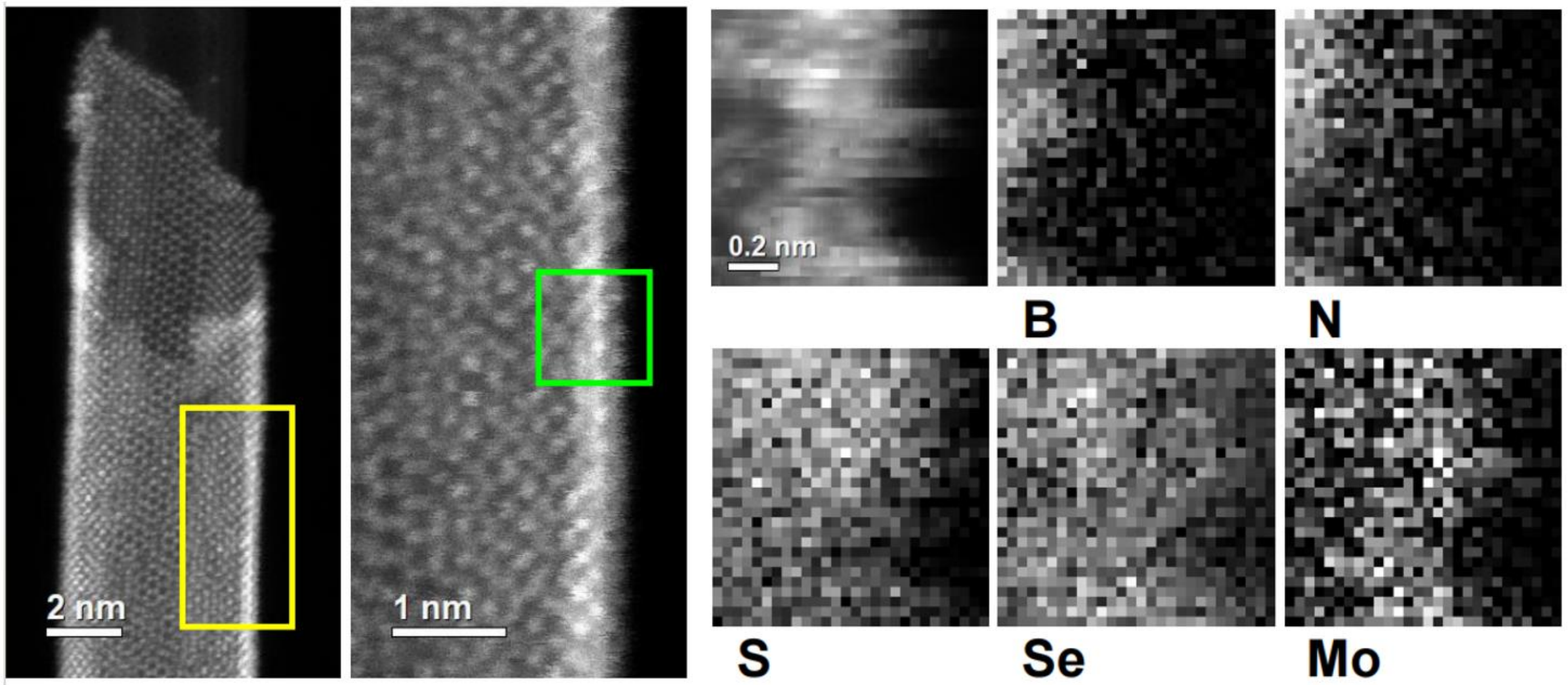

**Figure S3.** HAADF-STEM images of Janus MoSSe nanotubes and corresponding elements distribution of B, N, S, Se, Mo in the green frame area.

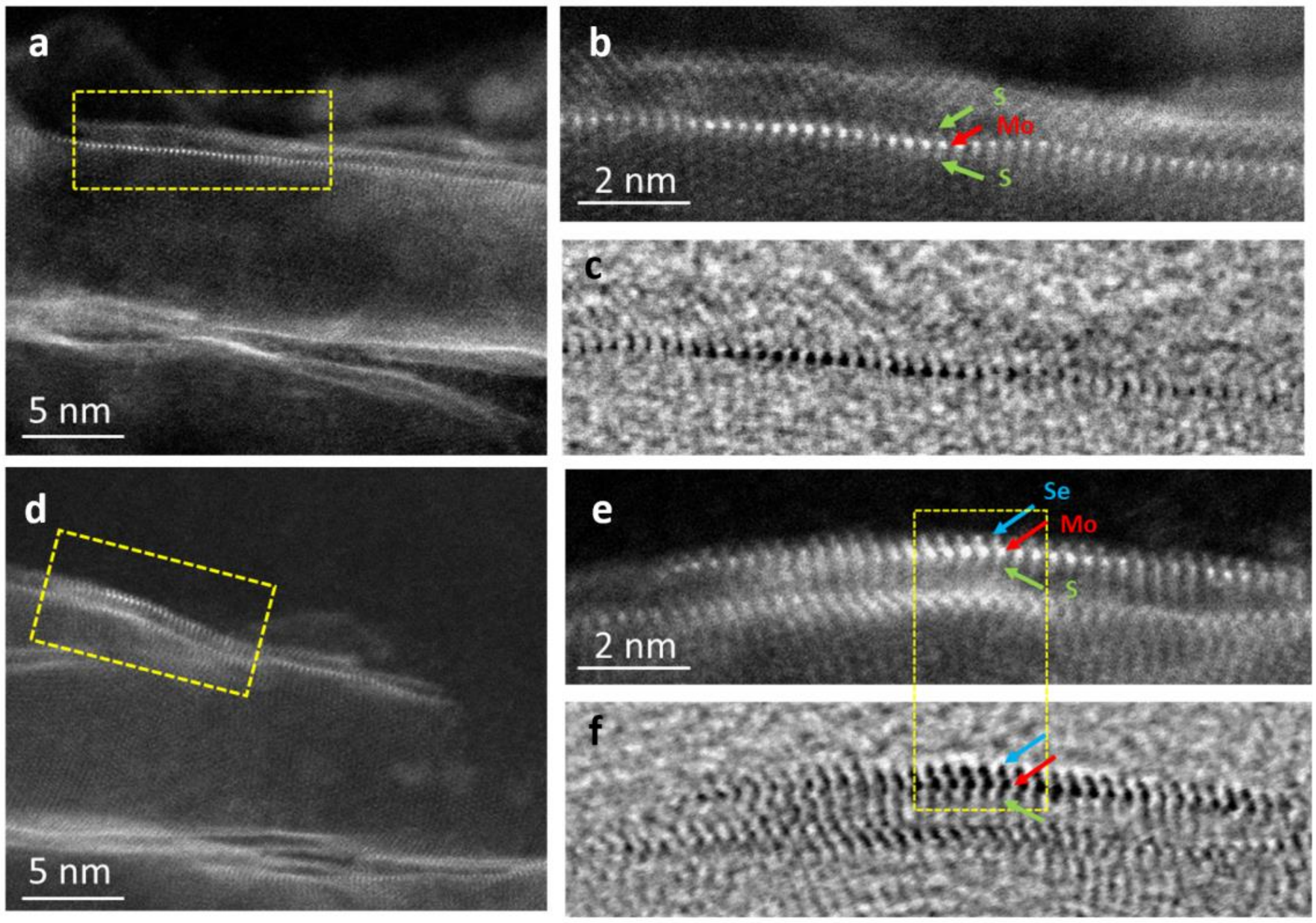

**Figure S4.** (a,d) HAADF-STEM images of the SWCNT-BNNT-TMDCNT. (b, c, e, f) is the magnified images in the corresponding yellow rectangle. S, Mo, and Se atoms are indicated by the green, red, and blue arrows.

In addition to the cross-section perpendicular to the axial direction of the nanotubes, some cross-sections parallel to the nanotubes have also been found, as shown in Figure S4. In the case of the nanotubes depicted in Figure S4a, a dual-layer of TMDC nanotubes wrapped on the SWCNT-BNNT template. For the inner layer of TMDC nanotubes, as evident from Figures S4b, c, the uniform and relatively dim brightness of both inner and outer atomic planes indicates that they remain as the pristine $MoS_2$ nanotube structure. However, for the nanotube samples presented in Figures S4d, f, the outermost TMDC nanotube exhibits a distinct appearance. Specifically, the intermediate Mo atomic layer is sandwiched between two distinctively bright chalcogen atomic layers, with the brighter outer layer attributable to Se atoms.

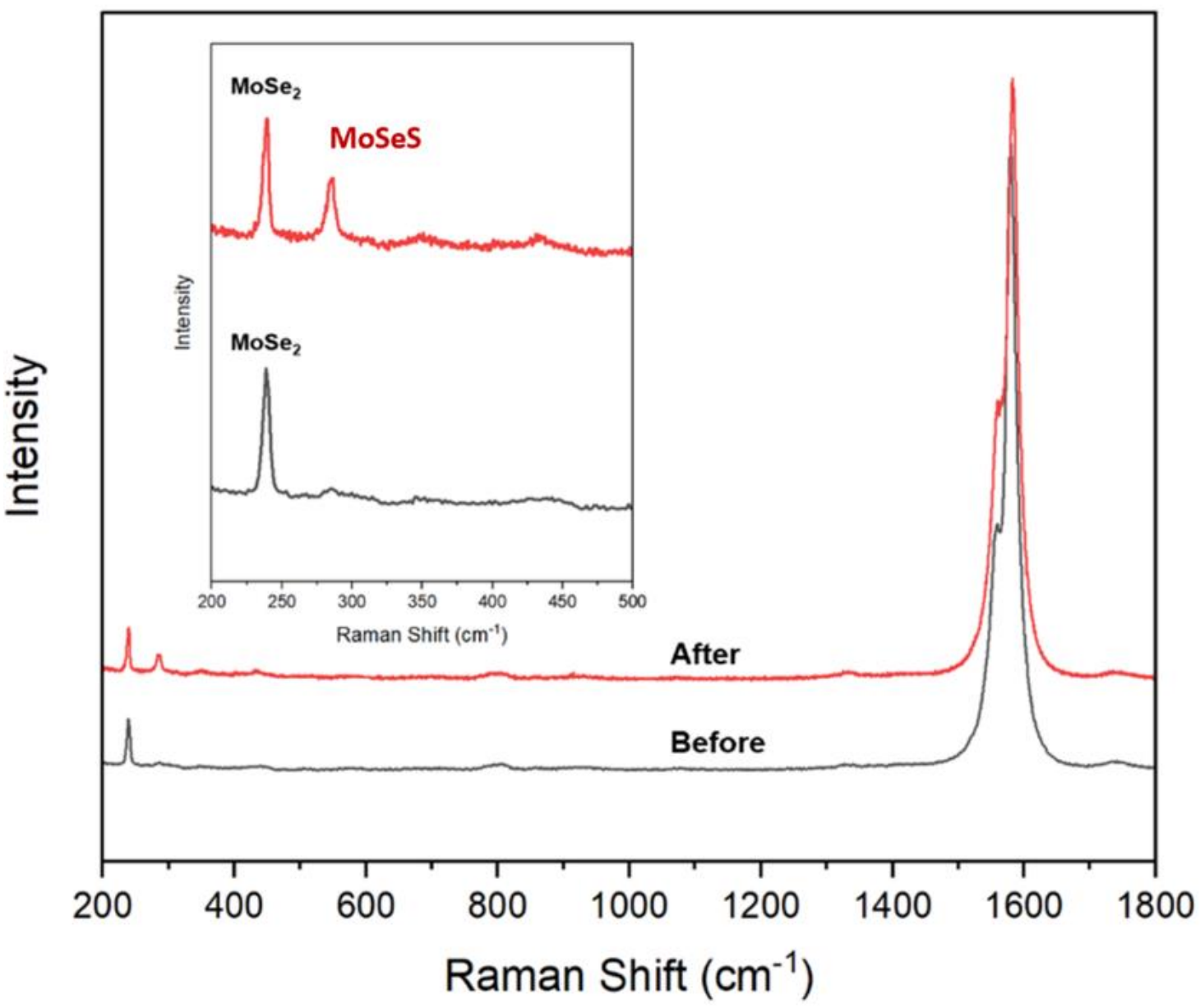

**Figure S5.** Raman spectra of coaxial heteronantubes after $MoSe_2$ synthesis (black line) and Janus MoSeS conversion processes (red line). The inset is the Raman spectra in the range of 200 to 500 $cm^{-1}$.

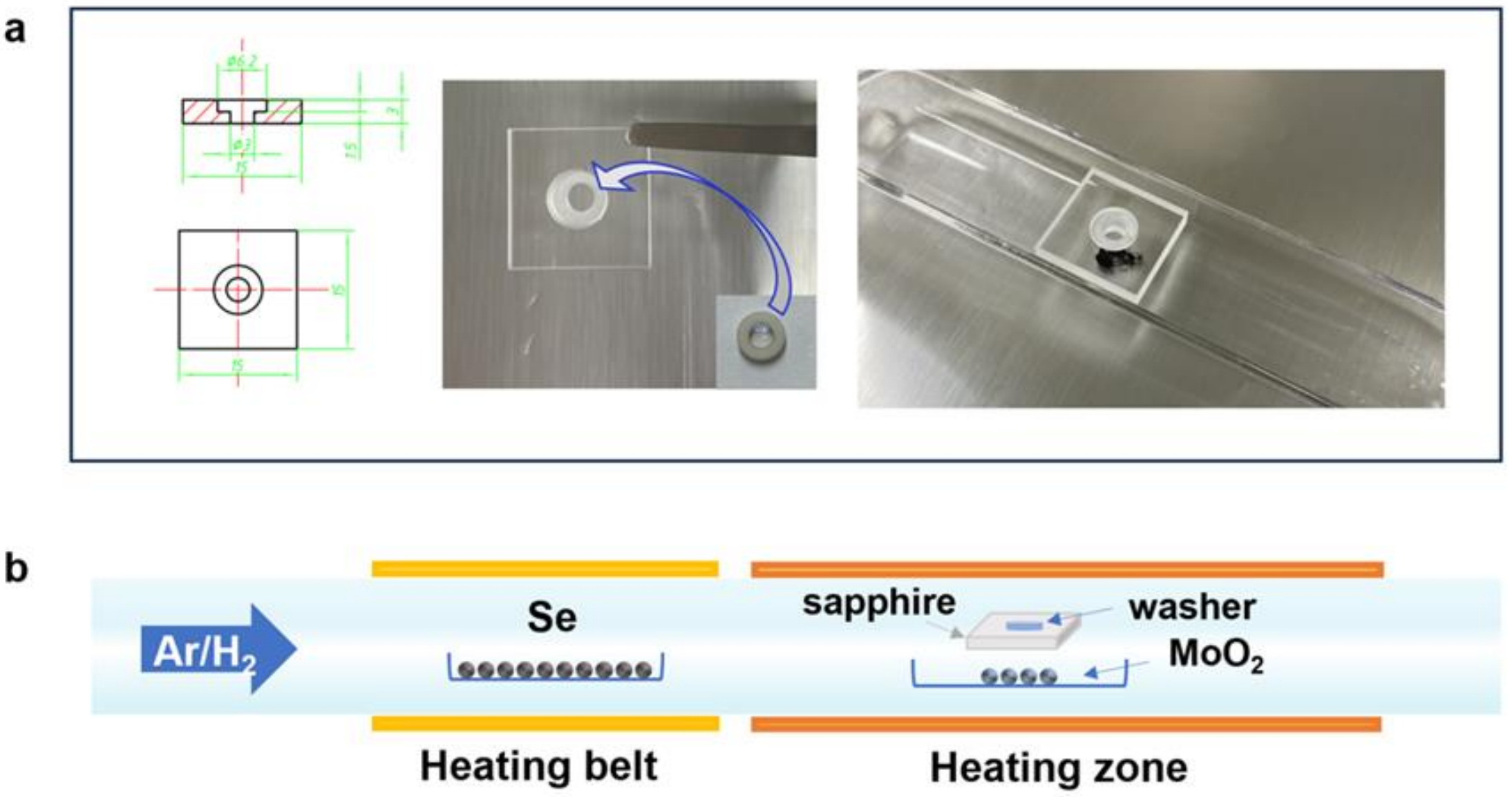

**Figure S6.** (a) Design drawing of sapphire support and schematic diagram of the placement of washer and support frame. (b) The schematic diagram of LPCVD growth of MoSe2 nanotubes on SWCNT-BNNT templates.